\documentstyle[12pt]{article}
\font\sqi=cmssq8
\def\DR{\rm I\kern-1.45pt\rm R}
\def\DC{\kern2pt {\hbox{\sqi I}}\kern-4.2pt\rm C}

\newcommand{\es}{\\[3mm]}
\makeatletter
\@addtoreset{equation}{section}
\makeatother

\advance\hoffset by -0.6cm
\begin{document}
\advance\voffset by -2cm
\centerline{ {\LARGE {\sc
            {\bf UNIVERSIT\'E DE GEN\`EVE}}}}
\centerline{ \raisebox{0mm}{{\small SCHOLA GEVENENSIS MDLIX}} }
\vspace{72mm}
\begin{figure}
\includegraphics{bigsigle.ps}
\end{figure}
\centerline{\LARGE On the Geometry of the}
\vspace{3mm}

\centerline{\LARGE Batalin-Vilkovisky Formalism}
%
%
\vspace{9mm}

\centerline{O. M. Khudaverdian$^1$\footnotetext{$^1$Supported in part
                          by the Swiss National Science Foundation.}  }
\vspace{2mm}

\centerline{{\small Yerevan State University }}
\centerline{{\small Yerevan (Armenia)  }}
\centerline{{\small{\it and} }}
\centerline{{\small D\'epartement de Physique Th\'eorique}}
\centerline{{\small Universit\'e de Gen\`eve}}
\centerline{{\small Gen\`eve  (Switzerland)}}
\vspace{10mm}

\centerline{A. P.
     Nersessian$^2$\footnotetext{$^2$E-MAIL:NERSESS@THEOR.JINRC.DUBNA.SU}}
\vspace{2mm}

\centerline{{\small Laboratory of Theoretical Physics}}
\centerline{{\small Joint Institute for Nuclear Research}}
\centerline{{\small Dubna, Head Post Office, P.O.Box 79,
                                         101 000 Moscow, Russia}}
\vspace{17mm}

\centerline{{\normalsize {\bf UGVA---DPT 1993/03--807}} }
\thispagestyle{empty}
\vfill
\pagebreak
\advance\voffset by 2cm
\null
\vspace{6cm}

\centerline{\Large{\bf Abstract}}\vspace{10mm}

\noindent An  invariant definition of the
operator $\Delta $ of the Batalin-Vilkovisky formalism
 is proposed. It is defined as the divergence of a Hamiltonian
 vector field with an odd Poisson bracket (antibracket).
Its main properties, which
 follow from this definition, as well as an example of  realization on
K\"ahlerian supermanifolds,  are considered. The geometrical meaning of
the Batalin-Vilkovisky formalism is discussed.

\setcounter{page}{0}
\thispagestyle{empty}

\vfill\pagebreak
\null
   \section{Introduction }
The Batalin--Vilkovisky formalism (BV--formalism) is the most
general method of quantization
of gauge field theories~\cite{bat1},~\cite{bat2}.
In recent years the interest
in studying  its geometrical nature has increased.
It was stimulated by  Witten's paper \cite{witten},
where the necessity of such an investigation is pointed out and
particularly for the
formulation of the background independent open-string
field theory on the base of the BV-formalism.
The realization of this program  began in \cite{witten1}, \cite{witten2}.

It is known that the BV-formalism uses unusual structures --  odd
Poisson brackets (antibrackets) and the operator $\Delta$.
As one of the main obstacles to the construction of
a background independent open
string field theory,  indicated by Witten \cite{witten},
 was the nonexistence  of an invariant
definition of the operator $\Delta$, and of a
naturally defined integral measure.
However, such a definition of the operator $\Delta$
was shown by one of us (O. K.) in
\cite{khud} before the cited Witten's paper  appeared (see also \cite{km}).
Its realization on K\"ahlerian supermanifolds  \cite{ners} and its
simplest properties
\cite{knjinr} were considered.

In the present paper we study the operator $\Delta$ in more detail.
We propose an invariant definition of this operator and show that
the condition of its nilpotency defines (in some sense) the choice of the
integration measure.

In Section 2 we propose an invariant definition of the operator $\Delta$
 on supermanifolds, given by the odd symplectic structure
and by the volume form as the divergence of the Hamiltonian vector field.
 We show that  all the relations  between the antibrackets and the operator
$\Delta$, which  are satisfied for canonical ones in the BV-formalism, are
satisfied for
any generalized operator $\Delta$  and the corresponding odd brackets.
However
the nilpotency condition for such an operator holds only for
a certain class of the integral density.

In Section 3 we consider the realization of the operator $\Delta $ on
supermanifolds, given by the odd (and even) K\"ahlerian structure,
and show that it is
nilpotent, if the integral density is the
characteristic class of the basic K\"ahlerian manifold (the function from
Chern classes). Then it corresponds to the
divergence operator  $\delta =* d *$
of the basic manifold.

In Section 3 we discuss the geometrical nature of
the Bat\-al\-in-Vil\-kov\-isky formalism.\\

When this paper was in preparation, we received two very important papers of
A.S. Schwarz \cite {schwarz}, \cite{schwarz2}  where the geometry of
the BV-formalism is analyzed in detail and in particular the same definition
of the operator $\Delta$, as in \cite{khud}, \cite{ners}, \cite{knjinr} and
in the present paper is given.
       \setcounter{equation}0
\section{Odd Poisson Brackets and Operator $\Delta$}
The  odd Poisson bracket (odd bracket, antibracket, Buttin bracket)
  of the functions $f$ and $g$ on the supermanifold ${\cal M}$ is
defined by the following  conditions \cite{ber}, \cite{leit} :
\begin{equation}
\begin{array}{l}
\{ f, ga +hb \}_{1} = \{ f, g\}_{1}a + \{ f, h \}_{1}b,
         \quad {\rm where}\quad a,b =  const   \es
p(\{ f, g \}_{1} )= p(f)+ p(g) + 1
         \quad {\rm (grading \ condition)}  \label{eq:bgrad} \es
\{ f, g \}_1 = -(-1)^{(p(f)+1)(p(g)+1)}\{ g, f \}_1
         \quad {\rm ( "antisymmetricity"\ condition )} \label{eq:anti} \es
\{ f, gh \}_1 =\{ f, g \}_{1}h +(-1)^{(p(f)+1)p(g)} g\{ f, h \}_1
         \quad {\rm ( Leibnitz \ rule )} \label{eq:bLieb} \es
 ( -1)^{(p(f)+1)(p(h)+1)}\{ f,\{ g, h \}_{1}\}_1
                  +{\rm {cycl. perm.}}( f,g,h) = 0
         \quad{\rm {(Jacobi\ id.)}}
\end{array}
\label{eq:bjac}
\end{equation}
  Locally, the  odd bracket can be written as:
\begin{equation}
\{ f, g \}_1 = \frac{\partial ^R f}{\partial x^A} \Omega^{AB}
            \frac{\partial  ^L g}{\partial x^B}
\label{eq:bloc}
\end{equation}
where $\Omega^{AB}$ satisfies to conditions
\begin{eqnarray}
& & p(\Omega^{AB} )= p(A)+ p(B) + 1
\quad  {\rm (grading\ condition)} \nonumber  \es
& & \Omega^{AB}  = -(-1)^{(p(A)+1)(p(B)+1)} \Omega^{BA}
\quad {\rm ( "antisymmetricity" \ condition )} \nonumber \es
& & ( -1)^{(p(A)+1)(p(C)+1)}\frac{\partial^{R}
 \Omega^{AB}}{\partial x^D }\Omega^{DC}
   +{\rm {cycl.\  perm.}}(A,B,C ) = 0
  \quad{\rm {(Jacobi\ id.)}}
\nonumber
\end{eqnarray}
where
$x^A$ are the local coordinates of ${\cal M}$, $p_A \equiv p( x^A ) $.
$ \frac{\partial ^R }{\partial x^A}$ and
$\frac{\partial ^L}{\partial x^A} $ denote
corresponding right and left derivatives.
They are connected with each other by:
$$\frac{\partial ^R f}{\partial x^A}=
  (-1)^{(p(f)+1)p_A}\frac{\partial ^l f}{\partial x^A}. $$
If ${\cal M}$ has an equal number of
even and odd coordinates, the odd bracket
can be nondegenerate. Then one can associate with it the
odd symplectic structure
\begin{equation}
\Omega =dx^A \Omega^{AB}dx^B   \label{eq:symp}
\end{equation}
 where $\Omega_{AB}\Omega^{BC} =\delta_{A}^{C}$.
This form is closed because of the Jacobi identities (\ref{eq:bjac}).
Locally, one can reduce  (\ref{eq:symp}) to the canonical form \cite{leit}:
\begin{equation}
\Omega^{\rm can} =\sum_{i=1}^{N} dx^i \wedge d \theta_i
   \label{eq:sympcan}
\end{equation}
where $( x^i,\theta _i) $ are some local coordinates (Darboux coordinates)
($p( x^i )= 0, p(\theta_i)= 1 $).
The corresponding odd bracket takes the form
\begin{equation}
		  \{f,g\}_1 =
			\sum_{i=1}^{N}\left(
		\frac{\partial^{R} f}{\partial  x^i}
		\frac{\partial^{L} g}{\partial\theta_i}
			 +
              \frac{\partial^{R} f}{\partial  \theta_i}
		\frac{\partial^{L} g}{\partial  x^i}
		   \right)       \label{eq:bcan}
\end{equation}
 Transformations preserving odd symplectic structures (or odd brackets)
 have (locally) the hamiltonian form:
  \begin{equation}
{\cal L}_{{\bf V}}\Omega = O
\quad {\rm iff}\quad {\bf V}=\{.,H\}_1 \equiv {\bf D}_{H}     \label{ham}
\end{equation}
 where $ H$ is an arbitrary function (Hamiltonian) on ${\cal M}$,
${\cal L}_{{\bf V}}$ denotes the Lie derivative
along the vector field {\bf V}.

It is well known that any supermanifold  can be associated with some vector
 bundle \cite{ber}. The odd symplectic structure
can be globally defined on the supermanifolds
 which are associated with the cotangent bundles of manifolds.

 Let $T^* M$ be the cotangent bundle of the manifold $M$.
  $x^{i}$ are local coordinates on $M$
  and $( x^{i}, v_{i} )$ are the corresponding
  local coordinates on $T^* M$.
 From map to map
  \begin{equation}
x^{i} \rightarrow {\tilde x^{i}} = {\tilde x^{i}} ( x),\quad
      v_{i} \rightarrow {\tilde v_{i}} = \sum_{i=1}^N
  \frac{\partial x^{j}}{\partial {\tilde x^{i}}} v_{j}.
     \label{eq:trans}
\end{equation}
 Considering for every map the superalgebra generated by
  $(x^{i}, \theta _{i})$, where the $x^{i}$ are even and
the  $\theta_{i}$ are odd coordinates,  transforming from map
 to map like $(x^{i}, v_{i})$ in ref{eq:trans})
  ($v\leftrightarrow \theta$), we come to the
 supermanifold ${\cal M}$ which is associated with $T^*M$ in the
  coordinates $(x^{i}, \theta_{i})$.

 Obviously, on this supermanifold, in the coordinates $(x^i, \theta_{i})$,
one can globally define the
 canonical odd symplectic structure (\ref{eq:sympcan}) \cite{leit}.

For the coordinates $(x^i, \theta_{i})$ on ${\cal M}$,
one can admit a more general class  of transformations:
                           $$
x^i\rightarrow {\tilde x}^i(x,\theta) \quad
	\theta_{i}\rightarrow {\tilde \theta}_{i}(x,\theta )    $$
   which do not correspond to (\ref{eq:trans}).
 In particular,
if $\theta_i \to {\tilde \theta}^i = \omega^{ij}\theta _j $,
  where $\omega _{ij}$ is the matrix of some nondegenerate Poisson bracket
 on $M$, then the supermanifold
  ${\cal M}$ in the coordinates $(x^i,{\tilde \theta}^i )$ is
  associated with the tangent bundle $TM$ of $M$,
e.g. ${\tilde \theta^i}$ transform
  under (\ref{eq:trans}) as $dx^i$.

 On the supermanifolds which can be associated in some
  coordinates with the tangent or cotangent bundle the
  superstructures are evidently reduced to  standard geometrical objects.

   Concerning the integration, the
properties of the odd brackets strongly
differ from the properties of odd brackets \cite {khud}, \cite {km},
such as:\\

 -- the odd bracket hasn't an invariant volume form and invariant
   integral densities;\\

 -- it has semidensities, which depend on higher order derivatives.\\

The first of this properties plays an essential role in
the Batalin--Vilkovisky quantization
formalism.
Using this property, we can construct on ${\cal M}$
an invariant generalization of the
important object of the BV-formalism  -- the operator $\Delta$.

Let the supermanifold ${\cal M}$ be
   provided with the odd symplectic structure (\ref{eq:symp})  and
 the volume form
\begin{equation}
dv=\rho(x, \theta)d^{N}x d^{N}\theta.  \label{eq:vol}
\end{equation}
Here $\rho (x,\theta)$
is some integral density. Under coordinate transformation
${\tilde x}^{A} = {\tilde x}^{A}(x)$,
it  transforms as:
\begin{equation}
   {\tilde \rho}({\tilde x}) = \rho (x ({\tilde x})) {\rm Ber}
 \frac{\partial^{R} x^A}{\partial {\tilde x}^B}  \label{eq:denstrans}
\end{equation}
On this supermanifold  one can invariantly define a second order odd
differential operator, which we call the "generalized operator $ \Delta $",
and which is invariant
 under the transformations preserving the symplectic structure
  and the volume form \cite {khud}. Its action on a function $f(x,\theta)$
  is the divergence  of the Hamiltonian
  vector field ${\bf D}_{f}$
  with the volume form $dv$:
\begin{equation}
\Delta _{\rho} f =\frac{1}{2}div_{\rho} {\bf D}_{f}
\equiv \frac{1}{2}\frac{{\cal L}_{{\bf D}_f} dv}{dv}, \label{eq:delta}
\end{equation}
where  ${\cal L}_{{\bf D}_f}$ is the Lie derivative
along ${{\bf D}_f}$ \cite {ber}, \cite {voronov}.
In coordinate form:
\begin{equation}
           \Delta f=\frac{1}{2\rho}
     \frac{\partial^R}{\partial x^A}
\left(\rho\{x^A,f\}_1\right)   \label{eq:deltaloc}
\end{equation}
   It has no analog within even symplectic structures.
The oddness of the Poisson bracket (\ref{eq:bloc})  forces
 the nontrivial grading of $\Delta $,
and the "antisymmetricity" condition (\ref{eq:anti}) forces
its dependence on second derivatives.

If the Poisson bracket  in (\ref{eq:delta}) is canonical,
and $\rho =constant$,
the generalized operator $\Delta$ takes the canonical  form
\begin{equation}
     \Delta^{\rm can} = \frac{\partial^R}{\partial x^i}
\frac{\partial ^L}{\partial \theta_i} \label{eq:deltacan}
\end{equation}
used in the BV-formalism.

 From the Leibnitz rule (\ref{eq:bLieb}) and the
definition (\ref{eq:delta}) follows
\begin{equation}
(-1)^{p(g)}\{f,g \}_1 = \Delta(fg) - f\Delta g
 -(-1)^{p(g)}(\Delta f)g  \label{eq:Liebdelta}
\end{equation}

 From the Jacobi identity (\ref{eq:bjac}) and the
definition (\ref{eq:delta}) follows
\begin{equation}
\Delta \{f,g \}_1 = \{f,\Delta g \}_1
+(-1)^{p(g)+1}\{\Delta f ,g \}_1     \label{eq:jacdelta}
\end{equation}
The density transformation rule (\ref{eq:denstrans}) implies for
the generalized operator $\Delta$
the following transformation rule
under canonical transformations:
\begin{equation}
 \Delta'f = \Delta f +\frac{1}{2}
\{\log {\cal J} ,f \}_1 , \label{eq:transdelta}
\end{equation}
where ${\cal J}$ is the Jacobian of the canonical transformation of
the odd bracket, $\Delta'$ is the
generalized operator $\Delta$ in the new coordinates.
For example, let us demonstrate the derivation of (\ref{eq:jacdelta}):
\begin{eqnarray}
&&\Delta \{f,g \}_{1}dv = {\cal L}_{{\bf D}_{\{f,g \}_1}}dv =\nonumber\es
&&=\left ( {\cal L}_{{\bf D}_f}{\cal L}_{{\bf D}_g} -
(-1)^{(p(f)+1)(p(g)+1)}{\cal L}_{{\bf D}_g}{\cal L}_{{\bf D}_f} \right )dv
= \nonumber\es
&&={\cal L}_{{\bf D}_f} {\Delta g}dv
-(-1)^{(p(f)+1)(p(g)+1)}{\cal L}_{{\bf D}_g} {\Delta f}dv = \nonumber\es
&& = \left ( \{f,\Delta g \}_{1}
+(-1)^{p(g)+1}\{\Delta f ,g \}_1 \right ) dv \nonumber
\end{eqnarray}
Let us write the following useful expressions, too:
\begin{equation}
 \Delta f(g)=f'(g) \Delta g + \frac{1}{2}f''(g)
\{g,g \}_1,  \label{eq:difdelta}
\end{equation}
 where $f(g)$ is an even complete function, and $g$ is
an even function on ${\cal M}$.

The properties (\ref{eq:Liebdelta}) -
(\ref{eq:difdelta}) are satisfied for any
$\rho$ and in the same manner as
the relations between canonical Poisson brackets
(\ref{eq:bcan}) and (\ref{eq:deltacan}) in the
BV-formalism \cite{bat2}, \cite{witten}.
This can br derived in the same way as (\ref{eq:jacdelta}).

However (\ref{eq:deltacan}) satisfies the
 nilpotency  condition
\begin{equation}
	       \Delta^2=0     \label{eq:nilp}
\end{equation}
which is very important in the BV-formalism.

The latter condition is violated for arbitrary $\rho (x,\theta)$.
 Indeed, if we have two densities $\rho$ and ${\tilde \rho}$,
and ${\tilde \rho} = \lambda\rho,\ p(\lambda )=0$,
then the corresponding operators $\Delta$ are related by
\begin{equation}
    \Delta_{\tilde \rho} f =\Delta_{\rho} f +
\frac{1}{2} \{\log \lambda, f\}_{1} \label{eq:deltacon}
\end{equation}
It is easy to see that
 \begin{equation}
 \Delta^{2}_{\tilde \rho} f =\Delta^{2}_{\rho} f
+ \{ \Gamma_{\lambda}, f \}_{1} ,  \label{eq:deltasqcon}
\end{equation}
where
\begin{equation}
  \Gamma_{\lambda}
=\lambda^{-\frac{1}{2}}\Delta_{\rho} \lambda^{\frac{1}{2}},
\quad p(\Gamma_{\lambda}) = 1
\end{equation}
If for some $\Delta_{\rho}$  the nilpotency condition
(\ref{eq:nilp}) is satisfied,
then it is also satisfied for  $\Delta_{\tilde \rho}$ (\ref{eq:deltacon})
if  $ \Gamma_{\lambda} =$odd constant$=0$.

For example, if the  symplectic structure is canonical,
(\ref{eq:nilp}) holds if
$\rho (x,\theta)$ satisfies the equation
\begin{equation}
\Delta^{{\rm can}} \sqrt \rho =0.\label{eq:sqrt}
\end{equation}
But this is the master equation of the
BV-formalism for the action $ S= -i\frac{1}{2}\log\rho $.

  The geometrical meaning of the
nilpotency condition (and correspondingly of the master equation)
will be illustrated on a simple example
in the next Section.
\setcounter{equation}0
\section{Example: The operator $\Delta$ on K\"ahlerian Supermanifolds}

As we saw in the previous section, in contrary to the case of an even
symplectic structure, on supermanifolds with an odd symplectic
stricture there arises
a nontrivial differential geometry.

 It is sufficient to show
the correspondence between the generalized operator
$ \Delta$ and geometrical objects on basic manifolds in the case of a
K\"ahlerian basic manifold,
because
on K\"ahlerian manifolds the symplectic structure corresponds to a
Riemannian one,
and a Riemmannian structure has a rich differential geometry.
Moreover, in this case there also  exists
 on ${\cal M}$ an even K\"ahlerian
structure, and, using it, we can construct a natural
integral density \cite {ners}, \cite{knjinr}.

Let ${\cal M}$ be a
complex supermanifold, and $z^A$ local complex coordinates
on ${\cal M}$.
A symplectic structure $\Omega^{\kappa}$ -- here and further
$\kappa =0(1)$ if
the symplectic structure is even (odd) --  on ${\cal M}$ is
called K\"ahlerian,
if in local coordinates $z^{A}$ it takes the following form:
\begin{equation}
\Omega^{\kappa}=i(-1)^{p(A)(p(B)+\kappa+1)}g^\kappa_{A {\bar B}}
           dz^A \wedge d{\bar z}^B,
\end{equation}
      where
$$  g^\kappa_{A {\bar B}}  =
     (-1)^{(p(A)+\kappa+1)(p(B)+\kappa+1)+\kappa +1}
         \overline {g^\kappa _{B {\bar A}}},
           \quad p(g^\kappa _{A\bar B})=p_A +p_B+\kappa$$

Then there exists a local real even (odd) function
$K^\kappa(z,{\bar z})$
    (K\"ahlerian potential), such that
\begin{equation}
	        g^\kappa_{A {\bar B}}  =
              \frac{\partial ^L}{\partial z^A}
		\frac{\partial ^R}{\partial {\bar z}^B}
			       K^\kappa  (z,{\bar z})
\end{equation}
         To  $\Omega^{\kappa}$ there corresponds
    the  Poisson bracket
   \begin{equation}
		     \{ f,g\}_\kappa
                        =
		     i\left(
	     \frac{\partial ^R f}{\partial \bar z^A}
			   g^{{\bar A}B}_\kappa
	        \frac{\partial  ^L g}{\partial z^B}
		           -
            (-1)^{(p(A)+\kappa)(p(B)+\kappa)}
      		 \frac{\partial  ^R f}{\partial z^A}
		     g^{{\bar A}B}_\kappa
		 \frac{\partial ^L g }{\partial \bar z^B}
		       \right),
\end{equation}
 where
$$g^{{\bar A}B}_\kappa g_{B{\bar C}}^\kappa=
  \delta^{\bar A}_{\bar C} \;\;,\;\;\;\; \overline{g^{{\bar A}B}_\kappa}
 = (-1)^{(p(A)+\kappa)(p(B)+\kappa)}g^{{\bar B}A}_\kappa.$$
 Its  satisfies the conditions of reality and "antisymmetricity"
       \begin{equation}
 \overline{\{ f, g\}_\kappa }
     =\{\bar f,\bar g \}_\kappa,\;\;\;\{ f, g \}_\kappa = -
(-1)^{(p(f)+\kappa)(p(g)+\kappa)}\{ g, f \}_\kappa,
\end{equation}
and the Jacobi identities :
\begin{equation}
( -1)^{(p(f)+\kappa)(p(h)+\kappa)}
\{ f,\{ g, h \}_{\kappa}\}_\kappa +{\rm {cycl. perm.}}(f,g,h) = 0
\end{equation}
Let ${\cal M}$ be associated with the tangent bundle $TM$ of the
K\"ahlerian manifold $M$,
and $z^A = (w^{a}, \theta^a )$  local coordinates on it,
$\theta^a$ transforming from map to map
like $dw^a$.
Let
\begin{equation}
g_{a{\bar b}}(w,{\bar w})
 =\frac{\partial^{2} K(w,{\bar w})}
{{\partial \omega^{a}}{\partial {\bar\omega^b}}}
\end{equation}
be a K\"ahlerian metric on $M$, with $K$ its K\"ahlerian
potential \cite{kobnom}.
Then  the local functions
\begin{eqnarray}
K_0(w,{\bar w},\sigma,{\bar \sigma})&=& K(w,{\bar w})+
ig_{a{\bar b}}(w,{\bar w})\sigma^a{\bar \sigma}^b,
\quad p( K_0 )=0 \label{eq:evenpot}\es
  K_{1}(w,{\bar w},\sigma,{\bar \sigma})&=&
    \epsilon\frac{\partial K(w,{\bar w})}{\partial w^a}\sigma^a+
  {\bar \epsilon}\frac{\partial K(w,{\bar w})}
   {\partial {\bar w^a}}{\bar \sigma^a} \quad p( K_1 )=1
 \label{eq:oddpot}
   \end{eqnarray}
(where  $\epsilon$ is
an arbitrary complex constant) correctly define an even and an
odd symplectic structures on ${\cal M}$
(this is not the most general form of K\"ahlerian potentials
on such supermanifolds \cite {ners})

The odd K\"ahlerian potential (\ref{eq:oddpot})
defines on ${\cal M}$ the following odd bracket:
\begin{equation}
  \{ f,g\}_1 = \frac{i}{\epsilon}\left (
	     \frac{\partial ^R f}{\partial \theta^a} \nabla^{a}g -
   \nabla^{a}f \frac{\partial^L g }{\partial
\theta^a } \right ) +{\rm c.c}  \label {eq:oddbk}
\end{equation}
where
\begin{equation}
{\overline {\nabla^a}}=g^{{\bar a}b}\nabla_b \quad \nabla_a=
\frac{\partial}{\partial w^a} -
        \Gamma^c_{ab}\theta^b\frac{\partial^L}{\partial\theta^c},
\end{equation}
and $\Gamma^c_{ab}=g^{\bar d c}g_{a \bar d,b} $ are
the Christoffel symbols of the K\"ahlerian metric  on $M$.

It is easy to see, that in the coordinates
$(w^a, \theta_a =i g_{a\bar b}{\bar \sigma}^b )$,
in which ${\cal M}$ is associated with $ T^* M$, the
odd Poisson bracket takes the canonical form.

The generalized operator $\Delta$ corresponding to (\ref{eq:oddbk}) takes the
form
    \begin{equation}
 \Delta f = \left(\frac{1}{\epsilon}\nabla^a
\frac{\partial^L}{\partial\theta^a} +
		\frac{1}{\bar \epsilon}
{\overline {\nabla^a}}\frac{\partial^L}{\partial{\bar \theta}^a}\right)f +
       \frac{1}{2}\{\log \rho,f \}_{1}
\end{equation}
If $\nabla_{a} \rho =0$ (or, in fact, if $\rho$ is a characteristic class
of $M$) then
   \begin{equation}
	    \Delta f = \frac{1}{\sqrt \rho}
  \left(\frac{1}{\epsilon}\nabla^a \frac{\partial^L}{\partial\theta^a} +
   \frac{1}{\bar \epsilon} {\overline {\nabla^a}}\frac{\partial^L}
    {\partial{\bar \theta}^a}\right)(\sqrt\rho f),  \label{eq:deltan}
\end{equation}
is obviously nilpotent.

The invariant density which corresponds to (\ref{eq:evenpot}) is
    \begin{equation}
	     \rho=  det(\delta^a_b+i
	      R^a_{bc{\bar d}}\theta^c \bar\theta^d)
\end{equation}
where $ R^a_{bc\bar d}=(\Gamma^a_{bc})_{,\bar d}$ is
the curvature tensor on $M$.
It is associated with the
generating functions of the Chern classes of the underlying
 K\"ahlerian manifold \cite{kobnom}.

Obviously (\ref {eq:deltan}) corresponds  to the operator
of covariant divergence $\delta = \ast d \ast$ on $M$ with some effective
weight.

 \setcounter{equation}0
\section { Discussion}

As we have seen in the previous Sections, the operator $\Delta$ has
a simple geometrical
nature on the  supermanifolds associated with the
cotangent bundles of manifolds.
Obviously, the same construction holds, if we replace the basic
manifold $M$ in the previous Sections by some
supermanifold ${\cal M}_{0}$. Then if $x^i$ are local coordinates on
${\cal M}_{0}$ ($p( x^i)\neq 0$), then, on the supermanifold ${\cal M}$ which
is associated with the cotangent bundle $T^*{\cal M}_{0}$,
one can naturally define
the odd Poisson bracket (\ref{eq:bcan})
 where $\theta_i$ corresponds to coordinates of the bundle, $p(\theta_i) =
 p( x^i ) +1$.
These coordinates are the  analogs of the antifields of the BV-formalism.

But what is the reason for the introduction of antifields
(and, correspondingly,
for the structure of supermanifolds with an odd symplectic structure)
 in the BV-formalism ?

In our opinion, this is connected to the
peculiarity of the integration on supermanifolds.
Indeed, if we have some differential form $\omega ( x^i, dx^i )$
on the supermanifold ${\cal M}_{0}$, its integral over ${\cal M}_{0}$
defined in the following way \cite{bernleit}, \cite{voronov}:
\begin{equation}
\int_{{\cal M}_{0}}\omega \equiv \int_{\hat{\cal M}_{0}}
\omega (x^i, \theta^i ) D(x,\theta)[dxd\theta]   \label{eq:blint}
\end{equation}
where $ p(\theta^{i})=p(x^{i}) + 1$, and $\hat{\cal M}_{0}$
denotes the supermanifold
associated with the tangent bundle of (supermanifold) ${\cal M}_{0}$
(i.e. $\theta^{i}$ transforms
like $dx^{i}$), then $D(x,\theta)$ is
the natural density on $\hat{\cal M}$ \cite{bernleit2}.

Transiting from the description on $\hat{\cal M}_{0}$ to that on
${\cal M} ={\cal T}_{*}{\cal M}_{0}$ --
the supermanifold associated with the cotangent bundle of
the supermanifold ${\cal M}_{0}$ -- we saw that
the integral (4.2) takes the form of the partition function of the
BV-formalism.

Correspondingly, the master-equation of the BV-formalism
corresponds to the closeness of the
initial differential form. This is clearly seen in the case where
the  basic manifold,
is K\"ahlerian (in the general case this proposition was
strongly proved in \cite{schwarz}.
Then the gauge invariance of the partition function in the BV-formalism
follows from  Stokes' theorem \cite{bernleit2}, \cite{voronov}.

\section {Acknowledgments}
We are very indebted to I. A. Batalin for valuable discussions.

\end{document}